# Variation of charge/orbital ordering in layered manganites $Pr_{1-x}Ca_{1+x}MnO_4$ investigated by transmission electron microscopy

X. Z. Yu [1, *], R. Mathieu [1], T. Arima [1, 2], Y. Kaneko [1, 3], J. P. He [1], M. Uchida [1], T. Asaka [4], T. Nagai [4], K. Kimoto [4], A. Asamitsu [1, 5], Y. Matsui [4] and Y. Tokura [1, 3, 6, 7]

[1] *Spin Superstructure Project, Exploratory Research for Advanced Technology, Japan Science and Technology Agency, Tsukuba, Ibaraki 305-8562, Japan*

[2] *Institute of Multidisciplinary Research for Advanced Materials, Tohoku University, Sendai 980-8577, Japan*

[3] *Multiferroics Project, Exploratory Research for Advanced Technology, Japan Science and Technology Agency, Tsukuba, Ibaraki 305-8562, Japan*

[4] *Advanced Electron Microscopy Group, National Institute for Materials Science, Tsukuba, Ibaraki 305-0044, Japan*

[5] *Cryogenic Research Center, University of Tokyo, Tokyo 113-8656, Japan*

[6] *Correlated Electron Research Center, National Institute of Advanced Industrial Science and Technology, Tsukuba, Ibaraki 305-8562, Japan*

[7] *Department of Applied Physics, University of Tokyo, Tokyo 113-8656, Japan*

(Dated: March 22, 2007)

Structural features of the charge/orbital ordering (CO/OO) in single-layered manganites $Pr_{1-x}Ca_{1+x}MnO_4$ ($0.3 \leq x \leq 0.65$) have been investigated systematically by transmission electron microscopy. Analyses of electron diffraction patterns as well as dark-field images have revealed that the CO/OO shows a striking asymmetric behavior as the hole doping $x$ deviates from $x = 0.5$. The modulation wavenumber linearly decreases with increasing $x$ in the over-hole-doped ($x > 0.5$) crystals, while much less dependent on $x$ in the under-hole-doped ($x < 0.5$) crystals. A temperature-induced incommensurate-commensurate crossover is observed in $0.35 < x < 0.5$ and $x = 0.65$. The correlation length of CO/OO in $x = 0.3$ was proven to become shorter than that in $0.35 \leq x \leq 0.65$.

Layered perovskite-type oxides have provided a good arena to study the quasi-two-dimensional charge dynamics in strongly correlated electron systems [1-8]. In $La_{1-x}Sr_{1+x}MnO_4$ ($x = 0.5$) [1, 7-8], which has been most frequently studied among all the single-layered manganites, the charge/orbital ordering (CO/OO) transition takes place at ~220 K. The modulation wavenumber $\delta$ is found to be commensurate with the lattice periodicity in $x = 0.5$. Research has also been extended to other La/Sr ratios [9-10]. In the over-doped ($x > 0.5$) compounds, the incommensurate $\delta$ value depends linearly on the hole doping $x$, while the commensurate CO/OO with $\delta$ of 1/2 is observed for $0.45 \leq x < 0.5$ at low temperatures. It was also reported [9] that only diffuse scattering exists in $x < 0.45$ due to the presence of disordered phase. The long-range magnetic order is limited in a narrow $x$-range near $x = 0.5$, in which the ferromagnetic zig-zag chains in a $MnO_2$ plane are antiferromagnetically coupled with each other (so-called CE-type magnetic structure). This long-range two-dimensional antiferromagnetic spin ordering is suppressed by random distribution of $e_g$ orbitals in $x < 0.45$. Recent experimental works have revealed the significant effect of quenched disorder on CO/OO even in the half-doped ($x$ = 0.5) single-layered manganites $RE_{0.5}Sr_{1.5}MnO_4$ (*RE*: rare-earth elements)[8]. The long-range CO/OO is suppressed by the quenched disorder which arises from the A-site (the trivalent rare-earth (*RE*) and divalent alkaline-earth (*AE*) ions) randomness. It can be measured by the variance in A-site ionic radii: $\sigma^2 = \Sigma_i x_i r_i^2 - r_A^2$. Here, $x_i$ and $r_i$ are the fractional occupancies and the effective ionic radii of A-site cation, respectively, and $r_A$ is the average radius of A-site. Since the variance in A-site ionic radii ($\sigma^2 \sim 2 \times 10^{-7}$ Å$^2$) for $Pr_{0.5}Ca_{1.5}MnO_4$ is much smaller than one for $La_{0.5}Sr_{1.5}MnO_4$ ($\sigma^2 \sim 1.7 \times 10^{-3}$ Å$^2$), the random distribution of $e_g$ orbitals due to the quenched disorder is almost avoidable in $Pr_{1-x}Ca_{1+x}MnO_4$ system. As a result, the stable long-range charge/orbital/spin ordering phase is expected to be created in this system. To reveal the detail structural features of the CO/OO in $Pr_{1-x}Ca_{1+x}MnO_4$, we have carried out a systematic study on $Pr_{1-x}Ca_{1+x}MnO_4$ ($0.3 \leq x \leq 0.65$) single crystals by means of transmission electron microscopy (TEM). We show the dependence of the modulation wave number and the microscopic CO/OO domain structure on the hole doping level including a temperature-induced incommensurate-commensurate (IC-C) crossover. The results are keys to explain the asymmetric behavior of magnetism in $Pr_{1-x}Ca_{1+x}MnO_4$.

Single crystals of $Pr_{1-x}Ca_{1+x}MnO_4$ ($0.3 \leq x \leq 0.65$) with various hole doping level $x$ have been grown by a floating-zone melt method. The phase purity and cation concentrations of $Pr_{1-x}Ca_{1+x}MnO_4$ ($0.3 \leq x \leq 0.65$) were checked by powder x-ray diffraction and inductively-coupled-plasma (ICP) atomic emission spectroscopy, respectively. Electron-transparent thin samples were prepared by mechanical polishing and subsequent argon-ion thinning with an acceleration voltage of 4 kV at room temperature. Selected-area electron diffraction (SAED) patterns and dark-field (DF) images were obtained by transmission electron microscopes, Hitachi HF-3000S and HF-3000L, both equipped with a cold-field emission gun and a liquid helium cooling holder. In order to determine the magnetic states and associated phase transition temperatures in $Pr_{1-x}Ca_{1+x}MnO_4$, ac-susceptibility was recorded as a function of temperature and frequency on a SQUID magnetometer (Quantum Design MPMSXL), which was equipped with the ultra low-field option and PPMS6000 [11].

First we focus on the half-doped ($x$ = 0.5) crystal. Figure 1(a) shows the [001]-zone SAED pattern at 20 K. The indices are based on an orthorhombic cell with $a$ = 5.41 Å, $b$ = 5.36 Å and $c$ = 10.73 Å. In addition to the fundamental spots, the SAED pattern reveals superlattice (SL) spots. It is well known that such SL spots should be attributed to the periodic arrangement of orbital stripes characterized by a modulation wave vector $\boldsymbol{q} = \delta \boldsymbol{a}^*$. Here, $\delta$ is the wavenumber corresponding to the inverse of the orbital stripe period. The SL spots vanish above ~325 K, which agrees with the CO/OO transition temperature $T_{CO/OO}$ determined by the previous resistivity measurements [11]. Analyses of the position of SL spots indicate that $\delta$ is commensurate ($\delta$ = 1/2) with the lattice periodicity at all the temperatures below $T_{CO/OO}$. Figure 1(b) shows a DF image obtained by using the SL spot (3/2, 2, 0) at 80 K. Bright areas represent CO/OO domains of a few hundred nanometers in size. In addition, many antiphase boundaries are observed as curved dark lines. The schematic of CO/OO in the half-doped sample is shown in Fig. 1(d). The $e_g$ orbitals, $d_{3x^2-r^2}$ and $d_{3y^2-r^2}$ of the $Mn^{3+}$ ions alternately arrange along the orthorhombic *a*-axis to form the $d_{3x^2-r^2}/d_{3y^2-r^2}$ type CO/OO or the orbital stripes. When the $d_{3x^2-r^2}$ orbital is replaced by the $d_{3y^2-r^2}$ orbital in the local structure or *vice versa*, the antiphase boundary for the ordered orbital stripes should appear. SL spots indicating the out-of-plane correlation can be also found at (n+1/2, 0, 2m+1) in the [010]-zone SAED, as shown in Fig. 1 (c), while they are much weaker than in the [001]-zone SAED. These spots which are newly observed in this single-layered manganite, may be ascribed to the CO/OO with ($\delta$, 0, 0) in the regularly tilted $MnO_6$ octahedra network with a modulation vector of (1 0 1) [12].

Next we show the hole doping $x$ dependence of CO/OO in $Pr_{1-x}Ca_{1+x}MnO_4$. Figure 2(a) presents the [001]-zone SAED patterns for various doping levels $x$ at 20 K. SL spots indicative of CO/OO appear in all the crystals with $0.3 \leq x \leq 0.65$. The intensity profiles of an $h$-scan presented in Fig. 2(b) clearly show that the modulation wavenumber $\delta$ varies with $x$: $\delta$ is larger than 1/2 for $x < 0.5$ and smaller than 1/2 for $x > 0.5$. The peak width $\gamma$ of the SL spot $(\delta, 2, 0)$ normalized by that of the fundamental spot $(0, 2, 0)$ is plotted against $x$ in Fig. 2(c). For $x$ = 0.65, $\gamma$ is close to 1.0, indicating the long-range CO/OO. As $x$ decreases from 0.65 to 0.35, $\gamma$ increases linearly and gradually. The coherent length of CO/OO along the $a$-axis in under-doped case becomes gradually shorter. For $x$ = 0.3, $\gamma$ suddenly increases to 2.0, indicating the relatively short-range CO/OO correlation in this crystal.

Change in the CO/OO domain structures with the hole doping is more clearly demonstrated by a DF images. Figure 3 shows DF images for $0.3 \leq x \leq 0.65$ at 80 K obtained by using the SL spot at $(\delta, 2, 0)$ in the corresponding SAED patterns shown in the insets. The images for $x > 0.3$ show the presence of stripe-shape bright/dark domains, the boundaries of which are found almost parallel to the <100>-axis in the tetragonal setting. These stripe-shape bright/dark domains were reversed in the DF image when another SL spot at $(2, \delta, 0)$ was used. Here one should note that the crystal structure of $Pr_{1-x}Ca_{1+x}MnO_4$ is orthorhombic even above $T_{CO/OO}$. In such orthorhombically distorted $K_2NiF_4$-type compounds, stripe-type domains are often induced by twins [8, 14] where the $a$ and $b$ axes alternate with each other. We can therefore conclude that these stripe-shape domains exhibit the CO/OO twin with two perpendicular directions, which should originate from the twin structures of the orthorhombic lattice distortion. The CO/OO domain size as well as the stripe-shape domain size decreases with the increase of twin. In the bright stripe-shape domains, one can also observe the random arrangement of smaller bright areas separated from each other by black lines (indicated by black arrows in Fig. 3(d)). A brighter area represents a single CO/OO domain. Black lines represent antiphase boundaries of the $e_g$ orbital ordering stripes as discussed in the half-doped case (see Fig. 1(b)). The average size of the single domains associated with the coherence of CO/OO also decreases with decreasing $x$. For the $x$ = 0.3 crystal, bright dots and dark areas are observed in DF image using the SL spot at $(\delta, 2, 0)$ (Fig. 3(a)). Each bright dot indicates the single CO/OO domain of nanometer size, much smaller than in $x$ = 0.65. In agreement with the SAED results in Fig. 2, the DF images show that the CO/OO domain size and hence the CO/OO correlation length decrease with increasing the nominal $e_g$-electron density (1-$x$).

The asymmetric doping behavior of the CO/OO with deviation of $x$ from 1/2 as well as the feature of the incommensurate-commensurate (IC-C) crossover is characterized as the change of the wavenumber $\delta$ with variations of temperature and hole doping level. Figure 4(a) shows the $T$ dependence of $\delta$ for various doping levels, as deduced from the peak position of a SL spot $(\delta, 2, 0)$ in the same condition. In the under-doped crystals ($0.3 \leq x < 0.5$), $\delta$ is larger than 1/2 and incommensurate with the lattice periodicity at low temperatures. As increasing $T$, $\delta$ merges to a commensurate value ($\delta$ = 1/2). The IC-C transition temperature ($T_{IC-C}$) is around 80 K for $x$ = 0.45, and around 240 K for $x$ = 0.40 and 0.35. Such a temperature-induced IC-C crossover phenomenon is not observed in $La_{1-x}Sr_{1+x}MnO_4$ system [9-10]. In the over-doped crystals (0.5 < $x$ < 0.65), $\delta$ is incommensurate with the lattice periodicity and independent of temperature. In $x$ = 0.65, $\delta$ is incommensurate ($\delta$ = 0.36) at 20 K, nearly satisfying the relation, $\delta$ = 1-$x$, and with warming above 200 K changes to a commensurate value ($\delta$ = 1/3). Figure 4(b) represents the hole doping $x$ dependence of $T_{CO/OO}$, the Neel transition temperature $T_N$ and the spin glass phase transition temperature $T_g$. $T_N$ and $T_g$ were determined by analyzing the ac-susceptibility curves [11]. In the over-doped region, the AFM phase as well as the long-range CO/OO was observed. In the under-doped region, the $T_{CO/OO}$ decreases with hole doping level $x$ and only short-range CO/OO as shown in Fig. 2 was observed in $x < 0.35$. This short-range structure may originate from the presence of extra $e_g$ electrons which also affects the exchange interaction. In fact, a spin glass state was observed in $x < 0.5$. This glassy short-ranged antiferromagnetic state is strongly correlated to the collapse of the long-range $e_g$ orbital ordering [11].

It is worth comparing the present results in $Pr_{1-x}Ca_{1+x}MnO_4$ with the case of pseudo-cubic $Pr_{1-x}Ca_xMnO_3$ [15-23]. Figure 5 shows the hole doping $x$ dependence of $\delta$. In the under-doped ($x < 0.5$) case, $\delta$ is dependent on $x$ at 20 K, while fixed at the commensurate value ($\delta \sim 1/2$) at 240 K. In the half- ($x = 0.5$) and over-doped ($x > 0.5$) crystals, $\delta$ conforms to $x$, obeying the relation that $\delta = 1-x$. The inset of Fig. 5 [16, 21-23] shows the relation between wavenumber $\delta$ and hole doping $x$ in $Pr_{1-x}Ca_xMnO_3$ at 20 K and 240 K. In the under-doped case ($x < 0.5$), a prototypical CO/OO pattern with $\delta = 1/2$ is observed. Extra electrons should occupy $Mn^{4+}$ chains along the $c$-axis [20]. The virtual hopping of $e_g$ electrons on $Mn^{4+}$ chains should cause ferromagnetic coupling along the $c$-axis, resulting in the canted antiferromagnetism at low temperature. In the under-doped single-layered compounds, however, $Mn^{4+}$ sites are isolated and do not form chains along the $c$-axis. Extra electrons on the $Mn^{4+}$ sublattices have no such energy gain due to the hopping along the $c$-axis as in the cubic case. This should lead to the formation of additional $Mn^{3+}$ stripes, and the unbalance between the numbers of $Mn^{3+}$ and $Mn^{4+}$ stripes should make the incommensurate CO/OO as observed. The short-range nature of CO/OO in $Pr_{1-x}Ca_{1+x}MnO_4$ with $x = 0.3$ may be also straightforwardly explained in terms of this scenario. In the half or over-doped $Pr_{1-x}Ca_xMnO_3$ crystals [19, 22], $\delta$ follows the relation that $\delta = 1-x$ below $T_N$, while the CO/OO with smaller $\delta$ values were observed above $T_N$. This implies that part of $e_g$ electrons, which form $Mn^{3+}$ stripes in the antiferromagnetic state, become itinerant with increasing $T$. The number of $Mn^{3+}$ stripes as well as $\delta$ should hence decrease in the paramagnetic phase. In the over-doped single-layered compounds with $0.50 < x \leq 0.60$, the $Mn^{3+}$ stripes are arranged as far apart from each other as possible even above $T_N$, which may be due to the strong localization of $e_g$ electrons. This type of CO/OO has been also observed in $La_{1-x}Sr_{1+x}MnO_4$ [9-10] and $Nd_{1-x}Ca_{1+x}MnO_4$ [3] systems and explained by the Wigner-crystal model [24]. The localization of $e_g$ electrons should become weaker as the doping concentration further increases, which may cause the IC-C crossover in $x = 0.65$ with increasing temperature. The origin of the IC-C phenomenon may be similar to the hole-doped $La_{2-x}Sr_xNiO_4$ [25] with $x \leq 1/3$, which has been interpreted as the entropy driven electron transfer between the on- and off-stripe regions.

In conclusion, structural features of the charge/orbital ordering (CO/OO) in $Pr_{1-x}Ca_{1+x}MnO_4$ ($0.3 \leq x \leq 0.65$) single crystals have been investigated systematically using the transmission-electron-microscopy. Compared with another canonical case of $La_{1-x}Sr_{1+x}MnO_4$, the long-range CO/OO phase exists over a wider range of the hole doping ($0.35 \leq x \leq 0.65$). Being different to the pseudo-cubic $RE_{1-x}AE_xMnO_3$ case, the modulation wave vector is unchanged between $T_N$ and $T_{CO/OO}$ in $x = 0.5$. As the hole doping $x$ deviates from 0.5, the electron-hole doping asymmetric behavior of the CO/OO shows up as the variation of the modulation wavenumber and the correlation length with the hole doping $x$. In the over-doped ($x > 0.5$) case, the CO/OO wavenumber depends linearly on $x$, suggesting the stripe-type ordering. In the under-doped ($x < 0.5$) crystals, the IC-C crossover was observed as increasing temperature. The real-space images show that large CO/OO domains in $x \geq 0.5$ turn into small nano-scale domains characteristic of the short-range one when $x = 0.3$. This asymmetric CO/OO behavior with the doping level $x$ is a key for understanding the asymmetric magnetic properties in this system.

We would like to thank C. Tsuruta for their technical assistance and Y. Tomioka and Y. S. Lee for valuable discussions. This work was partly supported by the Nanotechnology Support Project of Ministry of Education, Culture, Sports, Science and Technology (MEXT), Japan.

---

*Present address: Advanced Electron Microscopy Group, Advanced Nano Characterization Center, National Institute for Materials Science, 1-1 Namiki, Tsukuba, Postcode305-0044, Japan. Electronic address: Yu.xiuzhen@nims.go.jp.Tel: +81-29-851-3354 Ext. 8570, Fax: +81-29-860-4700.

Figure captions

Fig. 1: (a) [001]-zone selected-area electron diffraction (SAED) patterns at 20 K. The indices are based on the fundamental orthorhombic structure with the lattice parameters $a = 5.41$ Å, $b = 5.37$ Å, $c = 10.73$ Å (20 K). The SL spot (3/2, 2, 0) due to charge/orbital ordering (CO/OO) is indicated by the white arrow. (b) Dark field image obtained using the superlattice (SL) spot (3/2, 2, 0) at 80 K. The bright areas represent the CO/OO domains which vanish above the CO/OO transition temperature ($T_{CO/OO}$). (c) [010]-zone SAED patterns at 20 K. The CO/OO SL spot (3/2, 2, 3) is indicated by the white arrow. (d) Schematic of the CO/OO domains in the in-plane $Pr_{1-x}Ca_{1+x}MnO_4$ ($x = 0.5$) below $T_{CO/OO}$. The solid lines and dashed line show the CO/OO SL cell and an orbital antiphase boundary, respectively.

Fig. 2: (a) [001]-zone SAED patterns of $Pr_{1-x}Ca_{1+x}MnO_4$ ($0.3 \leq x \leq 0.65$) single crystals taken at 20 K. The indices of fundamental spots are based on the orthorhombic structure with the *Pccn* space group. The CO/OO SL spots ($\delta$, 2, 0) are indicated by white triangles. (b) Intensity profiles of the ($h$, 2, 0) scan for $x$ = 0.3, 0.4, 0.55 and 0.65 at 20 K. (c) Hole doping $x$ dependence of the peak width $\gamma$ for SL spot ($\delta$, 2, 0) normalized with that of the fundamental spot (0, 2, 0).

Fig. 3: Dark field images of $Pr_{1-x}Ca_{1+x}MnO_4$ crystals (a) $x = 0.3$, (b) 0.4, (c) 0.55, and (d) 0.65 at 80 K. These images are obtained by selecting the SL spot ($\delta$, 2, 0) indicated by white triangles. Bright areas denote the CO/OO domains corresponding to SL spots ($\delta$, 2, 0), which vanish above the CO/OO transition temperature ($T_{CO/OO}$). Insets show the corresponding SAED patterns.

Fig. 4: (Color online) (a) Temperature dependence of $\delta$ in $Pr_{1-x}Ca_{1+x}MnO_4$ ($0.3 \leq x \leq 0.65$). Dashed lines are the guide to the eyes. (b) Hole doping level $x$ dependence of $T_{CO/OO}$ (the transition temperature of CO/OO), T* (the transition temperature of short-range CO/OO), $T_N$ (Neel temperature) and $T_g$ (the transition temperature of spin glass state).

Fig. 5: (Color online) Hole doping level $x$ dependence of $\delta$ associated with the CO/OO at 20 K and 240 K. The inset as reproduced from ref. [16, 21-23] shows the $x$ dependence of $\delta$ in $Pr_{1-x}Ca_xMnO_3$ at 20 K and 240 K.

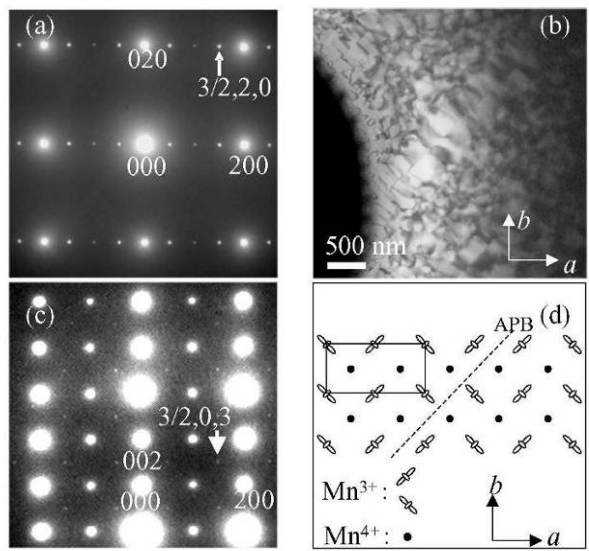

Fig. 1 Yu *et al.*

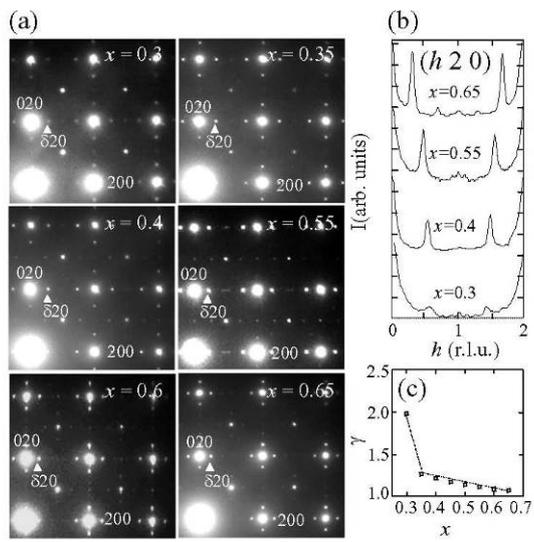

Fig. 2 Yu *et al*.

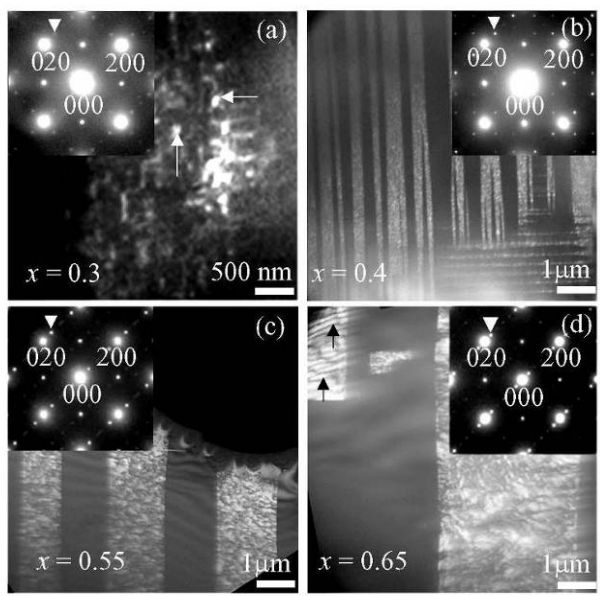

Fig. 3 Yu *et al*.

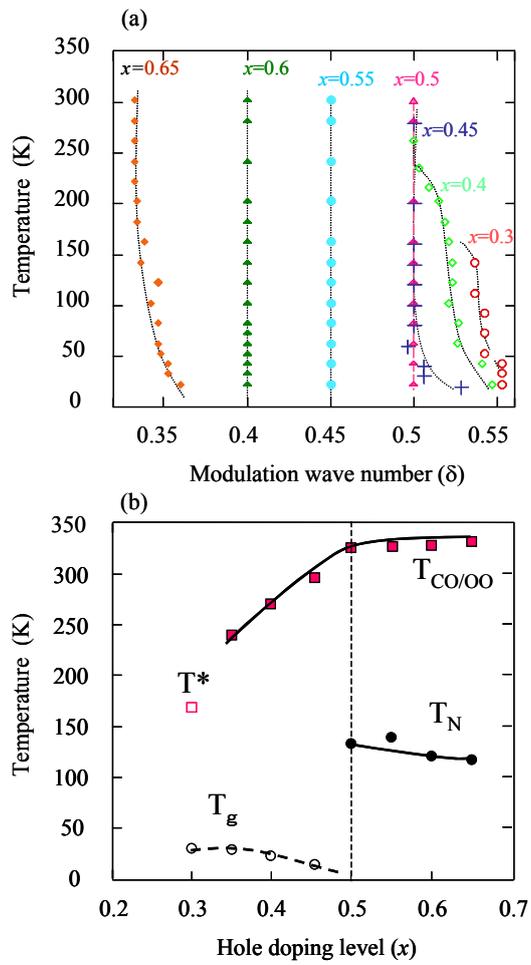

Fig. 4 Yu *et al.*

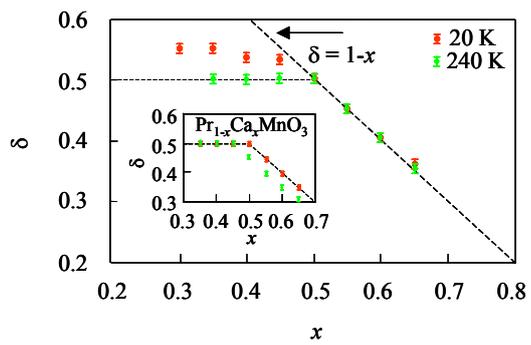

Fig. 5 Yu *et al*.

12